\magnification1095
\input amstex
\hsize32truecc
\vsize44truecc
\input amssym.def
\font\ninerm=plr9 at9truept
\font\twbf=cmbx12
\font\twrm=cmr12

\footline={\hss\ninerm\folio\hss}

\def\section#1{\goodbreak \vskip20pt plus5pt \noindent {\bf #1}\vglue4pt}
\def\eq#1 {\eqno(\text{\rm#1})}
\let\al\aligned
\let\eal\endaligned
\let\ealn\eqalignno
\let\dsl\displaylines

\let\o\overline

\let\a\alpha

\let\d\delta
\let\e\varepsilon
\let\f\varphi
\let\D\Delta

\let\G\varGamma
\let\P\varPsi
\let\F\varPhi

\let\z\zeta
\let\pa\partial
\let\t\widetilde
\let\u\tilde

\def\ns{\operatorname{ns}}
\def\sc{\operatorname{sc}}

\def\sn{\operatorname{sn}}
\def\cn{\operatorname{cn}}
\def\dn{\operatorname{dn}}
\def\sd{\operatorname{sd}}
\def\cd{\operatorname{cd}}
\def\nd{\operatorname{nd}}

\def\Re{\operatorname{Re}}

\def\ddt{\frac d{dt}}
\def\({\left(}\def\){\right)}
\def\[{\left[}\def\]{\right]}
\def\dr#1{_{\text{\rm#1}}}

\def\ct{constant}
\def\co{cosmological}

\def\il{inflation}

\def\pa{parameter}

\def\KK{Kaluza--Klein}
\def\JT{Jordan--Thiry}
\def\nos{nonsymmetric}
\def\up#1{\uppercase{#1}}
\def\eu{\expandafter\up}
\def\dint{-\kern-11pt\intop}

{\twbf
\advance\baselineskip4pt
\centerline{M. W. Kalinowski}
\centerline{\twrm (Higher Vocational State School in Che\l m, Poland)}
\centerline{On some \eu\co\ Consequences}
\centerline{of the Nonsymmetric}
\centerline{Kaluza--Klein (Jordan--Thiry) Theory}}

\vskip20pt plus5pt

{\bf Abstract.} In the paper we consider some consequences of the \eu\nos\ \KK
\ (\JT) Theory. We calculate primordial fluctuation spectrum functions, 
spectral indices, and first derivatives of spectral indices.

\vskip20pt plus5pt

In this paper we develop some ideas from Ref.~[1]. We will refer to Ref.~[1]
in all the details of our considerations.

Let us consider Eq.\ (3.313) from Ref.~[1].
$$
\varPhi(t)=\frac{(5+3a)\sqrt{5+4a}}{8\a_s(A_1\ns^2(u)+A_2)}
+\frac{\(1+\sqrt{5+4a}\)}{2\a_s} \eq1
$$
where
$$
\ealn{
u&=\root4\of{\frac{Ae^{n\varPsi_1}\(5Ae^{n\varPsi_1}-3\a_s^2r\)}{r^2}}
\cdot\sqrt{\sin\frac{\f+\pi}3}\,(t-t_0) &(2)\cr
A_1&=\sqrt{\frac53-a^2}\,\sin\frac{\f+\pi}3 &(3)\cr
A_2&=\sqrt{\frac53-a^2}\,\cos\frac{\f+4\pi}3+\frac1{12}(5+6a)&(4)
}
$$
and
$$
\ealn{
\cos\f&=-\frac{\sqrt A e^{\frac n2\varPsi_1}
\(9\a_s^2Br^2+40Ae^{n\varPsi_1}\)}{12\(5Ae^{n\varPsi_1}-3\a_s^2Br^2\)^{3/2}}
&(5)\cr
&0<a=\sqrt{\frac{\a_s^2Br^2}{Ae^{n\varPsi_1}}}<0.3115, &(6)
}
$$
$\ns$ is the Jacobi elliptic function for the modulus
$$
m^2=\frac{\sin\(\frac\f3\)}{\sin\(\frac{\f+2\pi}3\)}\,, \eqno(*)
$$
$B$ is an integration \ct, $r$ is a radius of manifold of vacuum states (a
scale of energy), $\a_s$ a coupling \ct. $A$ is given by the equation (3.164)
and is fixed by the details of the theory (see Ref.~[2]). $\P_1$~is a
critical value of a scalar field~$\P$ (see Ref.~[1]). $n$~is a dimension of
the group~$H$ (see Ref.~[1]). Eq.~(1) gives a solution to the equation of an
evolution of the Higgs field under some special assumptions (in the \co\
background). Due to this solution we are able to calculate $P_R(K)$---a
spectrum of primordial fluctuations in the Universe (see Ref.~[3]).

According to the Eq.~(3.29)
$$
P_R(K) = \(\frac{H_0}{2\pi}\)^2 \sum_{\u m,d}
\(\frac{H_0}{\ddt(\varPhi^d_{\u m})}\)_{\big| t=t^\ast}^2.
\eq7
$$
Using Eq.~(1) and the assumption
$$
\F^d_{\u m}(t)=\d^d_{\u m}\F(t) \eq8
$$
one gets for $t=t^*=\frac1{H_0}\ln\frac K{R_0H_0}$
$$
P_R(K)=P_0\,\frac{\(A_1+A_2\sn^2(v)\)^2\sc^2(v)}{\dn^2(v)} \eq9
$$
where $H_0$ is a Hubble \ct
$$
\ealn{
P_0&=\frac{12H_0^4 \a_s^2 n_1 A^{3/2}e^{(3/2)n\P_1}\cdot
\sin^{-3}\(\frac{\f+\pi}3\)} 
{\pi^2\(5\sqrt{Ae^{n\P_1}}+3\a_s\sqrt{Br^2}\)
\(5A\sqrt{Ae^{n\P_1}}+4\a_s\sqrt{Br^2}\)\(5Ae^{n\P_1}-3\a_s^2Br^2\)}\qquad &(10)\cr
v&=\(\ln K-t_0H_0-\ln(H_0R_0)\)\frac1{H_0}
\root4\of{\frac{Ae^{n\P_1}\(5Ae^{n\P_1}-3\a_s^2Br^2\)}{r^4}}
\sqrt{\sin\frac{\f+\pi}3}\,, &(11)
}
$$
$n_1$ means a dimension of the manifold $M=G/G_0$ (a vacuum states manifold),
$K^2={\vec K}^2$ means as usual the square of the length of a fluctuations
wave vector. We calculate a spectral index of a power spectrum
$$\ealn{
n_s(K)-1&=\frac{d\ln P_R(K)}{d\ln K}\,, &(12)\cr
n_s(K)-1&=2C\(\frac{\dn(v)}{\cn(v)\sn(v)}
-m^2\frac{\sn(v)\cn(v)}{\dn(v)}+ \frac{2A_2\sn(v)\cn(v)\dn(v)}
{A_1+A_2\sn^2(v)}\) &(13)
}
$$
and an important parameter
$$
\al
\frac{dn_s(K)}{d\ln K}&=2C^2\Biggl[
\(\sn^2(v)\dn(v)-\cn^2(v)\dn(v)-m^2\sn(v)\cn(v)\)\cr
&\qquad\qquad {}\times\(\frac1{\cn(v)\sn(v)}+\frac{m^2}{\dn(v)}\)\cr
&+2A_2\frac{\cn^2(v)\dn^2(v)-\sn^2(v)\dn^2(v)-m^2\sn^2(v)\cn^2(v)}
{A_1+A_2\sn^2(v)}\cr
&+4A_2^2\frac{\sn^2(v)\cn^2(v)\dn^2(v)}{\(A_1+A_2\sn^2(v)\)^2}\Biggr]
\eal
\eq14
$$
where $v$ is given by the formula (11) and all the Jacobi elliptic functions
in the formulae (9), (13), (14) have the same modulus
$$
m^2=\frac{\sin\frac\f3}{\sin\frac{\f+2\pi}3},
$$
$$
C=\frac1{H_0} \root4\of{\frac{Ae^{n\P_1}\(5Ae^{n\P_1}-3\a_s^2Br^2\)}{r^4}}
\sqrt{\sin\frac{\f+\pi}3}\,. \eq15
$$

In this theory we have some arbitrary parameters which can be translated to
parameters $A_1$, $A_2$, $C$ and~$m$. Thus we can consider Eqs (9), (13),
(14) as parametrized by $P_0$, $A_1$, $A_2$, $C$ and~$m$. Simultaneously $v$
can be rewritten as
$$
v=C\(\ln K-K_0'\) \eq16
$$
where $K_0'$ is an arbitrary parameter. Let us remind that in our theory we
have arbitrary \pa s: $r$, $\zeta$, $\xi$, $\a_s$ (they could be fixed by
some data from elementary particle physics, however, they are free as some
\pa s of the theory) and integration \ct s $B$ and~$t_0$ ($R_0$~also in some
sense). The interesting idea for future research consists in applying CMBFAST
software [4] to calculate all the \co\ CMB characteristics with mentioned \pa
s considered as free. Afterwards we should fit them to existing \co\ data~[5] as
good as possible in order to find some constraints for \pa s from our theory.

Let us consider the function $P_R(K)$ from Eq.~(9). It is easy to see that
the function
$$
f(v)=\frac{\(A_1+A_2\sn^2(v)\)^2 \sc^2(v)}{\dn^2(v)} \eq17
$$
is a periodic function
$$
f(v)=f(v+2K(m^2)l), \quad l=0,\pm1,\pm2,\ldots, \eq18
$$
$$
K(m^2)=\intop_0^{\pi/2} \frac{d\theta}{\sqrt{1-m^2\sin^2\theta}}\,. \eq19
$$
Thus we get
$$
\ealn{
P_R(K)&=P_R(Ke^{l\eta}), &(20)\cr
\eta&=\frac{2K(m^2)H_0\root4\of{\frac {r^4}{Ae^n\P_1\(5Ae^{n\P_1}-3\a_s^2Br^2\)}}}
{\sqrt{\sin\frac{\f+\pi}3}} &(21)
}
$$
or
$$
\eta=\frac{2K(m^2)}C. \eq22
$$
Thus it is enough to consider $P_R(K)$ in the interval
$$
e^{-\eta/2}\le K\le e^{+\eta/2} \eq23
$$
(a little similar to the first Brillouin zone). In order to have an
interesting \co\ region we should have
$$
\eta \simeq 20 \eq24
$$
which gives us a constraint
$$
\frac{2K(m^2)}C=\frac{2K(m^2)H_0}{\sqrt{\sin\frac{\f+\pi}3}}
\root4\of{\frac {r^4}{Ae^{n\P_1}\(5Ae^{n\P_1}-3\a_s^2Br^2\)}}
\approx10.\eq25
$$
Moreover we should consider a region with $n_s(K)\cong1$.
This can be satisfied if
$$
\al
&\dn^2(v)\(A_1+A_2\sn^2(v)\)-m^2\sn^2(v)\cn^2(v)\(A_1+A_2\sn^2(v)\)\cr
&\qquad{}+2A_2\sn^2(v)\cn^2(v)\dn^2(v)=0 
\eal
\eq26
$$
with the condition
$$
\sn(v)\cn(v)\dn(v)\(A_1+A_2\sn^2(v)\)\ne0. \eq27
$$

Using some simple relations among Jacobi elliptic functions and introducing a
new variable
$$
x=\sn^2(v) \eq28
$$
one gets a cubic equation
$$
x^3+\frac{A_1m^2-4A_2m^2-2A_2}{3A_2m^2}\,x^2
+\frac{3A_2-2A_1m^2}{3A_2m^2}\,x+\frac{A_1}{3A_2m^2}=0 \eq29
$$
or
$$
x^3+\frac{\rho m^2-4m^2-2}{3m^2}\,x^2+\frac{3-2\rho m^2}{3m^2}\,x+
\frac\rho{3m^2}=0, \eq30
$$
where
$$
\rho=\frac{A_1}{A_2}=\frac{\sqrt{\frac53-a^2}\,\sin\frac{\f+\pi}3}
{\sqrt{\frac53-a^2}\,\cos\frac{\f+4\pi}3+\frac{5+6a}{12}}\,. \eq31
$$

This equation can be reduced to
$$
y^3+py+q=0 \eq32
$$
where
$$
\ealn{
p&=\frac{-m^4(\rho^2+10\rho+16)+m^2(4\rho+11)-4}{27m^4} &(33)\cr
q&=\frac{2m^6(\rho^3+15\rho^2+92\rho-64)+6m^4(49+25\rho-2\rho^2)+6m^2
(43\rho-64)-16}{27^2m^6}\qquad &(34)\cr
x&=y-\(\frac{\rho m^2-4m^2-2}{9m^2}\). &(35)
}
$$
The discriminant of Eq.\ (32) can be written as
$$
D=\frac{q^2}{4}+\frac{p^3}{27} \eq36
$$
and one gets after some tedious algebra
$$
\al
D(\rho,m)=\frac1{27^4 m^{10}}\Bigl[
&-m^{10}\(387\rho^4+7376\rho^3+29456\rho^2+114944\rho+258048\)\cr
&{}+m^8\(243\rho^4+6660\rho^3+19062\rho^2+15000\rho+116352\)\cr
&{}+m^6\(246\rho^4+2292\rho^3-4716\rho^2+3299\rho+2473\)\cr
&{}+m^4\(144\rho^3+1512\rho^2+10476\rho+19251\)\cr
&{}+m^2\(-1548\rho^3+3584\rho^2+35190\rho-26460\)\cr
&{}+48\(75-39\rho\)\Bigr].
\eal
\eq37
$$

We use the Cardano formulae to find a root $x_0$. Afterwards we get
$$
v_\pm = \pm\intop_0^{\arcsin\sqrt{x_0}} \frac{d\theta}{\sqrt{1-m^2\sin^2
\theta}}\,. \eq38
$$
Thus we need $0\le x_0\le 1$. 

It is easy to see that every
$$
v_l=v_\pm+2K(m^2)l, \qquad l=0,\pm1,\pm2,\ldots, \eq39
$$
satisfies Eq.\ (26). 

If $D(m,\rho)>0$, we have only one real root
$$
x_0=\root3\of{-\frac q2+\sqrt D}+ \root3\of{-\frac q2-\sqrt D}
-\(\frac{\rho m^2-4m^2-2}{9m^2}\) \eq40
$$
where $q$ is given by Eq.\ (34) and $D$ by Eq.\ (37).
Moreover we need $0< x_0< 1$. In order to find a criterion let us transform
Eq.~(30) via
$$
x=\frac{1+w}{1-w}\,. \eq41
$$
This transformation maps a unit circle on a complex plane $|x|<1$ into $\Re
w<0$. One gets
$$
\al
&w^3+\frac{13m^2+\rho m^2-1+3\rho}{7m^2-3\rho m^2+5-\rho}\,w^2\cr
&\qquad{}+\frac{5m^2+3\rho m^2-5-3\rho}{7m^2-3\rho m^2+5-\rho}\,w
+\frac{\rho+1-m^2-\rho m^2}{7m^2-3\rho m^2+5-\rho}=0. 
\eal \eq42
$$
It is easy to see that the map (41) transforms roots of Eq.~(30) into roots
of Eq.~(42) and real roots into real roots. Especially roots inside the unit
circle into roots with negative real part.
Thus if Eq.~(30) possesses only one real root inside the unit circle then
Eq.~(42) possesses only one real root with the negative real part. Let $w_0$
be such a root of Eq.~(42) and $z$ and $\o z$ be the remaining roots. One gets
$$
\al
&(w-w_0)(w-z)(w-\o z)\cr
&\qquad{}=w^3-\(2\Re(z)+w_0\)w^2+
\(|z|^2+2\Re(z)w_0\)w-w_0|z|^2=0.
\eal \eq43
$$
Thus we only need 
$$
-w_0|z|^2>0, \eq44
$$
$$
x_0=\frac{1+w_0}{1-w_0}, \eq45
$$
which means
$$
\frac{\rho+1-m^2-\rho m^2}{7m^2-3\rho m^2+5-\rho}>0. \eq46
$$

The last condition means also that Eq.~(30) has a real root inside the unit
circle, i.e.~$x_0$. Now we only need to satisfy $x_0>0$, which simply means
$$
\frac \rho{3m^2}<0 \eq47
$$
or (if $m^2>0$)
$$
\rho<0. \eq48
$$
In this way we get conditions
$$
(\rho+1-m^2-\rho m^2)(7m^2-3\rho m^2+5-\rho)>0, \eq49
$$
$$
\rho<0. \eq50
$$

From (49--50) we get two solutions:
$$
\dsl{
\text{I} \hfill 0>\rho>-1,\quad 1>m^2>\frac{5-\rho}{7-3\rho} \hfill (51)\cr
\text{II} \hfill \rho<-1, \quad 0<m^2<\frac{5-\rho}{7-3\rho}\,. \hfill (52)}
$$
Simultaneously we have
$$
D(\rho,m)>0. \eq53
$$
Under these conditions our solution given by Eq.~(40) is non-negative with
modulus smaller than one.

In this way we have
$$
n_s(K)=1 \eq54
$$
for
$$
\ealn{
\ln K&=\frac{v_\pm+2K(m^2)l}{C}+K_0' &(55)\cr
K&=K_0 \exp\(\frac{v_\pm+2K(m^2)l}C\) &(56)
}
$$
where $K_0'=\ln K_0$.

Moreover we are interested in a deviation from a flat power spectrum. Thus we
consider
$$
n_s(K)=1\pm\frac{dn_s(K)}{d\ln K}\,\D\ln K \eq57
$$
where $\D\ln K\sim10$ and $\frac{dn_s(K)}{d\ln K}$ is calculated for $\ln K$
given by the formula (54). According to the formula (14) $\frac{dn_s(K)}{d\ln
K}$  is proportional to $C^2$.

Thus using (25) we get
$$
\frac{dn_s(K)}{d\ln K} \,\D\ln K \sim 10\(\frac{K(m^2)}{10}\)^2 \sim 0.1
\eq58
$$
which gives us rough estimate of a deviation from a flat power spectrum.

For $K$ given by Eq.\ (56) one gets
$$
\al
\frac{dn_s(K)}{d\ln K}&=2C^2 \biggl[\e\(\e_0x_0\sqrt{1-m^2x_0}
-\(1+x_0(m^2-1)\)\sqrt{1-x_0}\)\cr
&\qquad{}\times\frac{\sqrt{1-m^2x_0}+\e\e_0m^2\sqrt{x_0(1-x_0)}}{\sqrt{x_0(1-x_0)(1-m^2x_0)}}\cr
&+\frac{1+3m^2x_0^2-x_0(2m^2+1)}{\rho+x_0}+
\frac{4x_0(1-x_0)(1-m^2x_0)}{(\rho+x_0)^2}\biggr]
\eal
\eqno(*{*})
$$
where
$$
\e_0^2=\e^2=1. \eqno({*}{*}{*})
$$

It is interesting to estimate a range of $K$. One can consider
$$
K_0\exp\(\frac{v_+-K(m^2)}C\) \le K \le K_0 \exp\(\frac{v_++K(m^2)}C\)
\eqno({*}{*}{*}{*})
$$

Let us consider Eq.\ (57). If we find conditions for $\frac{dn_s(K)}{d\ln K}$
to be zero, we get a local flat spectrum (around $K$ given by Eq.~(56)). One
gets using (28) and $(*{*})$
$$
\al
&\(\e_0x_0\sqrt{1-m^2x_0}-\(1+x_0(m^2-1)\)\sqrt{1-x_0}\)\cr
&\qquad{}\times
\frac{\sqrt{1-m^2x_0}+\e\e_0m^2\sqrt{x_0(1-x_0)}}{\sqrt{x_0(1-x_0)(1-m^2x_0)}}\cr
&\quad{}=
\frac{x_0(2m^2+1)-1-3m^2x_0^2}{\rho+x_0}+
\frac{4x_0(x_0-1)(1-m^2x_0)}{(\rho+x_0)^2}\,. 
\eal
\eqno({*}{*}{*}{*}{*})
$$

For $x_0=x_0(m,\rho)$ (see Eq.\ (40)) is a known function of $m$ and~$\rho$,
the condition $({*}{*}{*}{*}{*})$ can be considered as a constraint for $m$ and~$\rho$.
Together with (51), (52) and~(53) it can give an interesting range for
parameters $m$ and~$\rho$. In this range the spectrum function (9) looks flat
(around $K$ given by Eq.~(56)).

Let us consider a formula for an evolution of Higgs' field given by Eq.\
(3.312) from Ref.~[1] corrected in the fifth point of Ref.~[2], (14.369):
$$
\varPhi(t)=\frac1{2\a_s}\(1\pm\sqrt{\frac5{1-e^{-5b(t-t_0)}}}\)
\eq59
$$
where $b=\frac{e^{n\P_1}A}{6r^2H_0}>0$,
$$
\lim_{t\to\infty}\F(t)=\frac1{\a_s}\(\frac{\pm\sqrt5+1}2\).
$$
It is easy to see that for
$$
\ealn{
t_{\max}-t_0&=\frac2b, &(61)\cr
\F\(\frac2b+t_0\)&\simeq \frac1{\a_s}\(\frac{1\pm\sqrt5}2\). &(62)
}
$$
Thus we can consider from practical point of view that an \il\ has been
completed for $t=t_0+\frac2b$. The \il\ starts for
$$
t\dr{initial}=t_0+\frac1{5b}\ln\frac54\,.\eq63
$$

Let us calculate $P_R(K)$, a spectral function for this kind of evolution of
Higgs' field. One gets from Eqs (7) and~(8)
$$
\ealn{
P_R(K)&=P_0\frac{\(\frac K{K_0}\)^{5{\bar\mu}}}{\(\(\frac
K{K_0}\)^{\bar\mu}-1\)^3}\,,&(64)\cr 
P_0&=\frac{4H_0^4\a_s^2n_1}{125\pi^2b^2} &(65)\cr
K_0&=R_0H_0e^{t_0H_0}. &(66)
}
$$
For a spectral index we get
$$
\ealn{
\frac{d\ln P_R(K)}{d\ln K}&=n_s(K)-1={\o\mu}\,\frac
{2\(\frac K{K_0}\)^{\bar\mu}-5}{\(\frac K{K_0}\)^{\bar\mu}-1} &(67)\cr
{\o\mu}&=\frac{5b}{H_0}. &(68)
}
$$
For the first derivative of $n_s$
$$
\frac{dn_s}{d\ln K}=\frac{8{\o\mu}^2\(\frac K{K_0}\)^{\bar\mu}}{\(\(\frac K{K_0}\)^{\bar\mu}
-1\)^2}\,. \eq69
$$

An interesting question is to find the range of $K$. Taking
$$
t\dr{initial}^\ast=\frac1{H_0}\(\frac{K\dr{initial}}{R_0H_0}\)=
t_0+\frac1{5b}\ln\frac54=t\dr{initial} \eq70
$$
one gets
$$
\ealn{
K\dr{initial}&=K_0\(\frac54\)^{1/{\bar\mu}} &(71)\cr
t^\ast_{\max}&=\frac1{H_0}\(\frac{K_{\max}}{R_0H_0}\)=t_0+\frac2b=t\dr{max} &(72)\cr
K_{\max}&=K_0 e^{2H_0/b}=K_0e^{10/{\bar\mu}} &(73)
}
$$
and finally
$$
K_0\(\frac54\)^{1/{\bar\mu}}<K<K_0 e^{10/{\bar\mu}}. \eq74
$$

Let us find $\t K$ such that
$$
n_s(\t K)=1. \eq75
$$
One gets
$$
\t K=K_0\(\frac52\)^{1/{\bar\mu}} \eq76
$$
$$
K_0\(\frac54\)^{1/{\bar\mu}}<\t K<K_0 e^{10/{\bar\mu}}. \eq77
$$
For such a $\t K$ one gets
$$
\ealn{
\frac{dn_s(\t K)}{d\ln K}&=5{\o\mu}^2 &(78)\cr
n_s(\t K)&\cong 1\pm 5{\o\mu}^2\D\ln K. &(79)
}
$$
If we take as usual $\D\ln K\simeq 10$, we get
$$
n_s(K)\cong 1\pm\frac{1250}{H_0^2}\,b^2=
1\pm\frac{625}{18r^4H_0^4}\,e^{2n\P_1}A^2. \eq80
$$
We can try to make it sufficiently closed to 1 taking a big $H_0$ (see
Eq.~(2.31) from Ref.~[1]).

In this case we have parametrized a spectral function and a spectral index by
only three parameters ${\o\mu}$, $P_0$ and~$K_0$, from which only ${\o\mu}$ matters.
Taking $K$ from the range
$$
1.2^{1/{\bar\mu}} < \frac K{K_0} < 10^{4.3/{\bar\mu}} \eq81
$$
we can calculate in principle all CMB characteristics (see Ref.~[4]) and to
fit to observational data (Ref.~[5]) (as good as possible) parameters ${\o\mu}$
(and~$K_0$). This will be a subject of future investigations.

Finally (in order to give a complete set of spectral functions) let us come
back to some formulae from Ref.~[1] (see Eqs (3.373--376)):
$$
\ealn{
P_R(K)&=P_0 g\(\frac K{R_0H_0}\) &(82)\cr
g(x)&=\frac1{x^2} \cdot \frac2
{\(\sd(u,\frac12)+\sqrt2 \cd(u,\frac12)\nd(u,\frac12)\)^2} 
\cr
\noalign{\eject}
u&=\sqrt2 C_1x+K\(\tfrac12\)-C_1\sqrt2 
-\intop_0^{\arccos(\sqrt5/C_1)}\frac{d\f}{\sqrt{1-\frac12\sin^2\f}}
&(83)\cr
P_0&=\(\frac{H_0}{2H}\)^2\cdot \frac{4\a_s^2n_1}{5C_1^2} &(84)\cr
n_s(K)-1&=-2-C_1x\cdot
\frac{\sqrt2 \cd(u,\frac12)-\sd^2(u,\frac12)}
{\sd(u,\frac12)+\sqrt2 \cd(u,\frac12)\nd(u,\frac12)} &(85)\cr
\frac{dn_s(K)}{d\ln K}&=-2C_1x \cdot
\frac{\sqrt2 \cd(u,\frac12)-\sd^2(u,\frac12)}
{\sd(u,\frac12)+\sqrt2 \cd(u,\frac12)\nd(u,\frac12)} 
-2C_1^2x^2\sd^2(u,\frac12)\cr
&-2C_1x^2\cdot
\frac{\(\sqrt2 \cd(u,\frac12)-\sd^3(u,\tfrac12)\)^2}
{\(\sd(u,\frac12)+\sqrt2 \cd(u,\frac12)\nd(u,\frac12)\)^2} &(86)
}
$$

In this case independent parameters are $C_1$ and $K_0=R_0H_0$. The amount of
\il\ in three cases can be calculated. In the first case (function (17)) it
is given by the corrected formulae (3.200--201) from Ref.~[1], i.e.
$$
\o N_0=H_0\sqrt{\frac {2r^2}{5Aae^{n\varPsi_1}}}
\intop_{\f_2}^{\f_1}\frac{d\theta}{\sqrt{1-\(\frac{4a+5}{10}\)\sin^2\theta}}
\eq87
$$
or
$$
\o N_0=\frac {H_0} {\sqrt{5\a_s}}\root4\of{\frac {4r^2}{ABe^{n\varPsi_1}}}
\intop_{\f_2}^{\f_1}\frac{d\theta}{\sqrt{1-\(\frac{4a+5}{10}\)\sin^2\theta}}
\eq88
$$
$$
\cos^2\f_1=\frac1{5+4a},\quad \cos^2\f_2=\frac5{5+4a}\,. \eq89
$$

It is interesting to notice that we can calculate $\o N_0$ from Eq.~(87) for
simplified Weinberg-Salam model (see Ref.~[6]) in the \nos\ version (see
Ref.~[2]). One gets
$$
\al
\o N_0&=J\cdot\sqrt{\frac2{15}}\(\frac{m_{\u A}}{m\dr{pl}\a_s}\)^{7/2}
2^{-27/4}\cr
&\times
\frac{\(7g(\z,\mu)+\sqrt{49g^2(\z,\mu)+384h(\z,\mu)}\)^{5/4}}
{\(2\mu^3+7\mu^2+5\mu+20\)^{7/4}|\z|^{1/4}(\mu^2+4)^{1/2}}\cr
&\times 
\frac{(1+\z^2)^{1/4}\(49g^2(\z,\mu)+g(\z,\mu)(\mu^2+4)
\sqrt{49g^2(\z,\mu)+384h(\z,\mu)}\)^{1/2}}
{\(\ln\(|\z|+\sqrt{\z^2+1}\)+2\z^2+1\)^{3/2}
\((5\z^2+4)\ln2-2\mu^2(1+\z^2)\)^{1/4}}
\eal
\eq87a
$$
where
$$
\ealn{
f(\z)&=\frac{16|\z|^3(\z^2+1)}{3(2\z^2+1)(1+\z^2)^{5/2}}
\(\z^2E\(\frac{|\z|}{\sqrt{\z^2+1}}\)-(2\z^2+1)K\(\frac{|\z|}{\sqrt{1+\z^2}}\)\)\cr
&+8\ln\(|\z|\sqrt{\z^2+1}\)
+\frac{4(1+9\z^2-8\z^4)|\z|^3}{3(1+\z^2)^{3/2}} &\text{(I)}\cr
g(\z,\mu)&=-f(\z)(\mu^2+4) &\text{(II)}\cr
h(\z,\mu)&=|\z|(2\mu^3+7\mu^2+5\mu+20)(1+\z^2)^{-1}\cr
&\times \((5\z^2+4)\ln2-2\mu^2(1+\z^2)\)
\(\ln\(|\z|+\sqrt{\z^2+1}\)+2\z^2+1\) &\text{(III)}\cr
J&=\intop_{\f_2}^{\f_1}\frac{d\theta}{\sqrt{1-\(\frac{4a+5}{10}\)\sin^2\theta}}
&\text{(IV)}\cr
a&=\(\frac{m\dr{pl}}{m_{\u A}}\)^7 2^{35/2}\a_s^{16}B^{1/2}\frac1{m_{\u A}}
\cr
&\times\(\frac{(2\mu^3+7\mu^2+5\mu+20)}
{(\mu^2+4)\(7g(\z,\mu)+\sqrt{49g^2(\z,\mu)+384h(\z,\mu)}\)}\)^{7/2}\cr
&\times \frac{(1+\z^2)^{1/2}\(\ln\(|\z|+\sqrt{\z^2+1}\)+2\z^2+1\)^4}
{|\z|^{1/2}\((5\z^2+4)\ln2 - 2\mu^2(1+\z^2)\)^{1/2}} &\text{(V)}
}
$$
and $a$ satisfies the condition (6), which is a condition for an integration
\ct~$B$, ${B>0}$. All of these formulae are subject to conditions
$$
\ealn{
\mu>\mu_0=-3.581552661\ldots \qquad&(\t R(\t\G)>0) &\text{(VIa)}\cr
|\z|>\z_0=1.36\dots \qquad&(\t P(\z)<0) &\text{(VIb)}\cr
\mu^2<\frac{(5\z^2+4)\ln2}{2(1+\z^2)} \qquad &(A>0). &\text{(VIc)}
}
$$

In the second case considered here the amount of an \il\ is infinite (in
principle). Moreover, it can be given by a finite formula
$$
\al
N&=H_0\(t\dr{end}-t\dr{initial}\)=H_0\(t\dr{max}-t\dr{initial}\)\cr
&=
H_0\(\frac2b-\frac1{5b}\ln\frac54\)=\frac{H_0}b\(2-\frac1{5}\ln\frac54\)
\cong 1.96\,\frac{H_0}b\,.
\eal
\eq90
$$

In the third case the amount of an \il\ is given by the formula (3.364) from
Ref.~[1]
$$
\sqrt2 C_2 - K\(\tfrac12\) = \intop_0^{\arccos\bigl(\frac1{2C_1e^{N_0}}\bigr)}
\frac{d\theta}{\sqrt{1-\frac12\sin^2\theta}} - \sqrt2 C_1 e^{N_0}, \eq91
$$
which is a transcendental equation for $N_0$.

All of those $P_R(K)$ functions will be treated by numerical methods from
Ref.~[4] and afterwards we compare the results with observational data (e.g.\
Ref.~[5]). This will be the first stage of an application of our theory in
cosmology. Afterwards we try to estimate intrinsic parameters of our theory
and to start a multicomponent \il.

\def\ii#1 {\item{[#1]}}
\section{References}
\setbox0=\hbox{[11]\enspace}
\parindent\wd0

\ii1 {Kalinowski} M. W., 
{\it Dynamics of Higgs' field and a quintessence
in the nonsymmetric Kaluza--Klein (Jordan--Thiry) theory}, 
arXiv: hep-th/0306241v1. 

\ii2 {Kalinowski} M. W., 
{\it Nonsymmetric Fields Theory and its Applications\/},
World Scientific, Singapore, New Jersey, London, Hong Kong 1990.

\item{} {Kalinowski} M. W., 
{\it Nonsymmetric Kaluza--Klein (Jordan--Thiry) Theory in a general
nonabelian case}\/,
Int. Journal of Theor. Phys. {\bf30}, p.~281 (1991).

\item{} {Kalinowski} M. W., 
{\it Nonsymmetric Kaluza--Klein (Jordan--Thiry) Theory in the electromagnetic
case}\/,
Int. Journal of Theor. Phys. {\bf31}. p.~611 (1992).

\item{} {Kalinowski} M. W., 
{\it Can we get confinement from extra dimensions}\/,
in: Physics of Elementary Interactions (ed. Z.~Ajduk, S.~Pokorski,
A.~K.~Wr\'oblewski), World Scientific, Singapore, New
Jersey, London, Hong Kong 1991.

\item{} {Kalinowski} M. W., 
{\it Scalar fields in the nonsymmetric Kaluza--Klein (Jordan--Thiry)
theory}, arXiv: hep-th/0307242.

\ii3 {Liddle} A. R., {Lyth} D. H., 
{\it Cosmological Inflation and Large-Scale Structure}\/,
Cambridge Univ. Press, Cambridge 2000.

\ii4 {\tt www.cmbfast.org}

\ii5 {Bennett} C. L. et al., 
{\it First Year Wilkinson Microwave Anisotropy Probe (WMAP) Observations:
Determination of Cosmological Parameters}\/,
astro-ph/0302209v2.

\ii6 {Manton} N. S.,
{\it A new six-dimensional approach to the Weinberg-Salam model\/},
Nucl. Phys. {\bf B158}, p.~141 (1979).

\end